\documentclass[
preprint,
nobibnotes,
amsmath,amssymb,
 aps,
]{revtex4-1}

\usepackage{graphicx}
\usepackage{dcolumn}
\usepackage{bm}

\begin{document}

\preprint{APS/123-QED}

\title{Bond-specific reaction kinetics during the oxidation of (111) Si: Effect of n-type doping\\}

 \author{B. Gokce}

\author{D.E. Aspnes}

 \author{G. Lucovsky}

 \author{K. Gundogdu}
 \affiliation{%
 Physics Department, North Carolina State University, Raleigh NC 27695 \\}%

\date{\today}

\begin{abstract}
It is known that a higher concentration of free carriers leads to a higher oxide growth rate in the thermal oxidation of silicon. However, the role of electrons and holes in oxidation chemistry is not clear. Here, we report real-time second-harmonic-generation data on the oxidation of H-terminated (111)Si that reveal that  high concentrations of electrons increase the chemical reactivity of the outer-layer Si-Si back bonds relative to the Si-H up bonds.  However, the thicknesses of the natural oxides of all samples stabilize near 1 nm at room temperature, regardless of the chemical kinetics of the different bonds.
\begin{description}
\item[Usage]
Secondary publications and information retrieval purposes.
\item[PACS numbers]
\end{description}
\end{abstract}

\pacs{Valid PACS appear here}
\maketitle

The passivation of Si substrates and the chemical stability of passivated surfaces are both significant for Si-based device technology.\cite{Mizsei} The most common way to passivate silicon surfaces is to terminate them with H by dipping them into HF or NH${_4}$F solutions.  This process strips the oxide layer and caps the outermost orbitals of the Si substrate with H. 

The chemical stability of these surfaces depends on the reactivity of two types of bonds: the outermost Si-H bonds and the Si-Si back bonds to the underlying layer. Determining the individual chemical reactivities of these bonds is essential for understanding and controlling chemical processes on H-terminated silicon surfaces. Under ambient conditions the dominant chemical process is oxidation.  The oxidation rate depends on environmental conditions such as temperature, pressure, humidity, and the possible presence of reacting agents as well as intrinsic sample properties such as surface morphology and type and density of free carriers.\cite{Kakiuchi,Yamada,Ligenza} The effect of carrier concentration on oxidation at elevated temperatures has been characterized by spectroscopic ellisometry (SE), \cite{Irene,Fuoss} surface differential reflectance spectroscopy (SDR)\cite{Takizawa}, Auger electron spectroscopy (AES)\cite{Kamiura}, and X-ray photoelectron spectroscopy (XPS)\cite{Ying}, all of which probe oxide thickness. The oxidation rate has been found to increase with increasing carrier concentration.

The studies cited above determine the macroscopic behavior of oxidation, but do not provide information about the reactivities of the individual types of bonds. Here, we report results obtained by second-harmonic-generation anisotropy (SHGA) and SE measurements that address the chemical reactivities of the different types of bonds that are present in H-terminated (111) Si. The bond sensitivity of SHGA is apparent when the data are analyzed using the anisotropic bond model (ABM)\cite{Adles,Powell} of nonlinear optics. This approach shows that the radiated SHG signal is maximized when a particular bond is aligned parallel to the driving field, i.e., when the acceleration of that bond charge is itself maximized.  As samples are rotated, the back bonds realize this condition sequentially, allowing the respective hyperpolarizabilities to be extracted from SHGA data.\cite{Gokce} Therefore, changes in the chemical nature of these bond are easily detected by SHGA. The SHGA response of the "up" bonds is isotropic, allowing their chemical reactivity to be inferred from the signal offset. Our results show that under ambient laboratory conditions the reactivity of the first-layer Si-Si bonds is significantly affected by carrier concentration. Nevertheless, despite these differences in early oxidation rates, overall oxide thicknesses evolve similarly and stabilize near 1.0 nm for all doping levels. 

The data reported here were obtained on 10$^{12}$, 10$^{14}$, and 10$^{16}$ cm$^{-3}$ P-doped (111)Si wafers and one 10$^{18}$cm$^{-3}$ As-doped (111) Si wafer, all purchased from Virginia Semiconductor. The samples were cleaned by consecutive 10-min immersions in 80 $^{\circ}$C NaOH/H$_{2}$O$_{2}$/H$_{2}$O (1:1:5) and 80 ¡C HCl/H$_{2}$O$_{2}$/H$_{2}$O (1:1:5) solutions.  Native oxides were then stripped and the samples capped with H by a 20-min immersion in $40\%$ NH$_{4}$F. To prevent pitting, the NH$_{4}$F solution was deoxygenated prior to immersion.\cite{Wade} Measurements began approximately 90 s after NH$_{4}$F immersion, after the surfaces were dried with high-purity N$_{2}$. 

SHGA data were obtained in a reflection geometry using p-polarized light at an angle of incidence of 45$^{\circ}$ for both illumination and detection.  The exciting beam was generated by a Ti-sapphire oscillator.  It consisted of 100 fs pulses centered at 806 nm, arriving at a repetition rate of 70 MHz. The SHG response was detected by a photomultiplier every 1$^{\circ}$  as the sample was rotated by 360$^{\circ}$. Further details are provided elsewhere.\cite{Gokce} 

Figures 1a, b, and c show the data for the P-doped samples. All are normalized to the level where the SHG responses no longer evolve with time. Where present, the three major peaks correspond to the near alignment of one of the three back bonds to the direction of the exciting field. As expected, the overall effect of oxidation is to increase the magnitude of the SHG signal at these azimuths, a consequence of the higher electronegativity of O relative to H. The evolution clearly depends significantly on carrier concentration.  For the lowest two carrier concentration the SHG intensity initially increases then decreases to its terminal value, whereas for the most heavily doped sample the SHG intensity increases monotonically. This striking dependence on carrier concentration is even more evident in Fig. 1d, which shows the average evolution of the signal at these three azimuths for all four samples, including that of the As-doped sample for which n =10$^{18}$cm$^{-3}$. To relate these differences to the chemical kinetics of the bonds, we determined the respective hyperpolarizabilities by fitting the data to the predictions of the ABM using expressions given in [10]. The hyperpolarizability results are summarized in Fig. 2. 

The effect of chemistry on hyperpolarizabilities can be understood as follows. Generation of SHG by a bond requires that the bond be asymmetric. While an Si-H bond is obviously asymmetric, it is less obvious that terminating an up bond with H generates contributions from the Si-Si back bonds as well. The reason is chemical induction.  The greater electronegativity of H removes charge from the Si atom to which it is attached, making the Si-Si back bonds asymmetric as well. Thus the SHG contribution of the back bonds tends to track that of the up bond.  If the top H is now replaced by OH, the greater electronegativity of O results in greater asymmetries and hence stronger signals for both up and back bonds. This type of response is clearly evident in Fig. 2 for the lightly doped sample. 

With this background, the striking difference among the samples of different carrier concentrations can now be explained. In all cases, oxidation begins with the replacement of H with OH, as described above. The next step is the insertion of the O into a back bond, leaving the surface still capped with H.  For lightly doped samples this step is slow. But for heavily n-type samples the reactivity of the back bonds is much higher, and replacement and insertion occur at similar rates.  Thus the sequential increase and decrease in SHG responses seen for the relatively lightly doped samples becomes a monotonic increase. The same behavior is seen for both P and As doping, indicating that the effect is due to carrier concentration, not type of dopand. Experiments on p-type substrates show additional differences, illustrating that carrier type is also important. These  results will be presented elsewhere. 

Possible complications include bulk-quadrupole and electric-field-induced-second-harmonic (EFISH) contributions.\cite{Hirayama} However, any bulk contribution will show no time dependence and hence, even if present, can be neglected. EFISH contributions are isotropic, result from charge trapping in the overlayer, and can only be a factor for heavily doped material.\cite{Xu} We investigated EFISH for our samples by illuminating them in a glove box that was pressurized with N$_{2}$ to prevent oxidation.  For H termination the SHG intensity did not increase, as expected.  For heavily doped oxidized samples the signal increases by about 15\% before saturating. Thus EFISH is not a factor, and the hyperpolarizabilities that we determine are an accurate representation of the oxidation chemistry of the different bonds.

To investigate whether this difference extends beyond the top two Si layers, we obtained thickness data on the same set of H-terminated samples as a function of time using SE.  Details are provided elsewhere.\cite{Aspnes}  Given the demonstrated dependence of oxidation kinetics on doping, the results are surprising. Except for an initial transient, Fig. 3 shows that the oxide thicknesses increase at a common rate up to about 1 nm, when the measurements were terminated. The final data in Fig. 3 include an approximately 0.2 nm contribution from adsorbed hydrocarbons.  In fact weeks after these measurements the thicknesses remains almost constant. This is consistent with other data that we have obtained, including those of a sample that we have measured occasionally for over 30 years. These results indicate that even if the initial chemical reactivities of the different types of bonds are significantly different, this distinction disappears after about a monolayer or so of oxidation. 

The limiting ca. 1 nm oxide thickness can be understood in terms of the electronic coherence length for medium range order (MRO) in noncrystalline SiO$_2$. \cite{Lucovsky} MRO extends beyond the first- and second-nearest-neighbor distances of $\sim$0.16 to $\sim$0.31 nm, respectively. As deduced from the position of the first sharp diffraction peak (FSDP) in the Fourier transform of X-ray diffraction data, MRO includes third- and fourth-neighbor pair correlations extending to 0.5 nm, as well as rings of bonded atoms. A coherence length of $\sim$1 nm, obtained from the width of the FSDP, has been interpreted in terms of rigid 6-member rings encapsulated by more compliant 5- and 7-member rings.\cite{Du} This defines a characteristic cluster size that is effectively-strain-free. These clusters are connected through the more compliant lower-symmetry rings and any further degree of longer range order, e.g., nanocrystalline periodicity, is not present. Given that the free surface provides no constraints, oxidation can then be expected to proceed up to a thickness where the encapsulated 6-member rings are formed, at which point additional energy would be required for further oxide growth.  We thus conclude that the formation of a native oxide on Si is strain-driven to a thickness of $\sim$1 nm, after which it displays long-term stability measured in years. This process is not dependent on the particular growth surface, as evidenced by the symmetry of the SHG response. 

In conclusion, our data show a distinct carrier-concentration dependence of the initial oxidation process of H-terminated (111)Si. Oxidation of the outer-layer Si-Si back bonds is significantly enhanced with strong n-type doping. To our knowledge this is the first report of the dependence on carrier concentration of the chemical reactivity of different classes of bonds of this technologically important material. It is also the first report to recognize that while chemical reactivities of bonds are doping-dependent, the terminal thickness of the SiO$_2$ overlayer is independent of carrier concentration, and is determined independently by the MRO length scale of SiO$_2$.
\newpage

\bibliography{APLbib1}

\providecommand{\noopsort}[1]{}\providecommand{\singleletter}[1]{#1}%
\begin{thebibliography}{18}%
\makeatletter
\providecommand \@ifxundefined [1]{%
 \@ifx{#1\undefined}
}%
\providecommand \@ifnum [1]{%
 \ifnum #1\expandafter \@firstoftwo
 \else \expandafter \@secondoftwo
 \fi
}%
\providecommand \@ifx [1]{%
 \ifx #1\expandafter \@firstoftwo
 \else \expandafter \@secondoftwo
 \fi
}%
\providecommand \natexlab [1]{#1}%
\providecommand \enquote  [1]{``#1''}%
\providecommand \bibnamefont  [1]{#1}%
\providecommand \bibfnamefont [1]{#1}%
\providecommand \citenamefont [1]{#1}%
\providecommand \href@noop [0]{\@secondoftwo}%
\providecommand \href [0]{\begingroup \@sanitize@url \@href}%
\providecommand \@href[1]{\@@startlink{#1}\@@href}%
\providecommand \@@href[1]{\endgroup#1\@@endlink}%
\providecommand \@sanitize@url [0]{\catcode `\\12\catcode `\$12\catcode
  `\&12\catcode `\#12\catcode `\^12\catcode `\_12\catcode `\%12\relax}%
\providecommand \@@startlink[1]{}%
\providecommand \@@endlink[0]{}%
\providecommand \url  [0]{\begingroup\@sanitize@url \@url }%
\providecommand \@url [1]{\endgroup\@href {#1}{\urlprefix }}%
\providecommand \urlprefix  [0]{URL }%
\providecommand \Eprint [0]{\href }%
\@ifxundefined \urlstyle {%
  \providecommand \doi  [0]{\begingroup \@sanitize@url \@doi}%
  \providecommand \@doi [1]{\endgroup \@@startlink {\doibase
  #1}doi:\discretionary {}{}{}#1\@@endlink }%
}{%
  \providecommand \doi  [0]{doi:\discretionary{}{}{}\begingroup
  \urlstyle{rm}\Url }%
}%
\providecommand \doibase [0]{http://dx.doi.org/}%
\providecommand \Doi [0]{\begingroup \@sanitize@url \@Doi }%
\providecommand \@Doi  [1]{\endgroup\@@startlink{\doibase#1}\@@Doi}%
\providecommand \@@Doi [1]{#1\@@endlink}%
\providecommand \selectlanguage [0]{\@gobble}%
\providecommand \bibinfo  [0]{\@secondoftwo}%
\providecommand \bibfield  [0]{\@secondoftwo}%
\providecommand \translation [1]{[#1]}%
\providecommand \BibitemOpen [0]{}%
\providecommand \bibitemStop [0]{}%
\providecommand \bibitemNoStop [0]{.\EOS\space}%
\providecommand \EOS [0]{\spacefactor3000\relax}%
\providecommand \BibitemShut  [1]{\csname bibitem#1\endcsname}%
\bibitem [{\citenamefont {Mizsei}(2002)}]{Mizsei}%
  \BibitemOpen
  \bibfield  {author} {\bibinfo {author} {\bibfnamefont {J.}~\bibnamefont
  {Mizsei}},\ }\href@noop {} {\bibfield  {journal} {\bibinfo  {journal}
  {Vacuum},\ }\textbf {\bibinfo {volume} {67}},\ \bibinfo {pages} {59}
  (\bibinfo {year} {2002})}\BibitemShut {NoStop}%
\bibitem [{\citenamefont {{H. Kakiuchi}}\ \emph {et~al.}(2007)\citenamefont
  {{H. Kakiuchi}}, \citenamefont {{H. Ohmi}}, \citenamefont {{M. Harada}},
  \citenamefont {{H. Watanabe}},\ and\ \citenamefont {{K.
  Yasutake}}}]{Kakiuchi}%
  \BibitemOpen
  \bibfield  {author} {\bibinfo {author} {\bibnamefont {{H. Kakiuchi}}},
  \bibinfo {author} {\bibnamefont {{H. Ohmi}}}, \bibinfo {author} {\bibnamefont
  {{M. Harada}}}, \bibinfo {author} {\bibnamefont {{H. Watanabe}}}, \ and\
  \bibinfo {author} {\bibnamefont {{K. Yasutake}}},\ }\href@noop {} {\bibfield
  {journal} {\bibinfo  {journal} {App. Phys. Lett.},\ }\textbf {\bibinfo
  {volume} {90}},\ \bibinfo {pages} {091909} (\bibinfo {year}
  {2007})}\BibitemShut {NoStop}%
\bibitem [{\citenamefont {Yamada}(1999)}]{Yamada}%
  \BibitemOpen
  \bibfield  {author} {\bibinfo {author} {\bibfnamefont {H.}~\bibnamefont
  {Yamada}},\ }\href@noop {} {\bibfield  {journal} {\bibinfo  {journal} {J.
  Appl. Phys},\ }\textbf {\bibinfo {volume} {86}},\ \bibinfo {pages} {5968}
  (\bibinfo {year} {1999})}\BibitemShut {NoStop}%
\bibitem [{\citenamefont {Ligenza}(1961)}]{Ligenza}%
  \BibitemOpen
  \bibfield  {author} {\bibinfo {author} {\bibfnamefont {J.~R.}\ \bibnamefont
  {Ligenza}},\ }\href@noop {} {\bibfield  {journal} {\bibinfo  {journal} {J.\
  Phys.\ Chem.},\ }\textbf {\bibinfo {volume} {65}},\ \bibinfo {pages} {2011}
  (\bibinfo {year} {1961})}\BibitemShut {NoStop}%
\bibitem [{\citenamefont {Irene}\ and\ \citenamefont {Dong}(1978)}]{Irene}%
  \BibitemOpen
  \bibfield  {author} {\bibinfo {author} {\bibfnamefont {E.~A.}\ \bibnamefont
  {Irene}}\ and\ \bibinfo {author} {\bibfnamefont {D.~W.}\ \bibnamefont
  {Dong}},\ }\href@noop {} {\bibfield  {journal} {\bibinfo  {journal} {J.
  Electrochem. Soc.},\ }\textbf {\bibinfo {volume} {125}},\ \bibinfo {pages}
  {1146} (\bibinfo {year} {1978})}\BibitemShut {NoStop}%
\bibitem [{\citenamefont {Fuoss}\ and\ \citenamefont {Topich}(1980)}]{Fuoss}%
  \BibitemOpen
  \bibfield  {author} {\bibinfo {author} {\bibfnamefont {D.}~\bibnamefont
  {Fuoss}}\ and\ \bibinfo {author} {\bibfnamefont {J.~A.}\ \bibnamefont
  {Topich}},\ }\href@noop {} {\bibfield  {journal} {\bibinfo  {journal} {Appl.
  Phys. Lett.},\ }\textbf {\bibinfo {volume} {36}},\ \bibinfo {pages} {275}
  (\bibinfo {year} {1980})}\BibitemShut {NoStop}%
\bibitem [{\citenamefont {{J. Takizawa}}\ \emph {et~al.}(2006)\citenamefont
  {{J. Takizawa}}, \citenamefont {{S. Ohno}}, \citenamefont {{J. Koizumi}},
  \citenamefont {{K. Shudo}},\ and\ \citenamefont {{M. Tanaka}}}]{Takizawa}%
  \BibitemOpen
  \bibfield  {author} {\bibinfo {author} {\bibnamefont {{J. Takizawa}}},
  \bibinfo {author} {\bibnamefont {{S. Ohno}}}, \bibinfo {author} {\bibnamefont
  {{J. Koizumi}}}, \bibinfo {author} {\bibnamefont {{K. Shudo}}}, \ and\
  \bibinfo {author} {\bibnamefont {{M. Tanaka}}},\ }\href@noop {} {\bibfield
  {journal} {\bibinfo  {journal} {J. Phys.: Condens. Matter},\ }\textbf
  {\bibinfo {volume} {18}},\ \bibinfo {pages} {L209} (\bibinfo {year}
  {2006})}\BibitemShut {NoStop}%
\bibitem [{\citenamefont {Kamiura}\ \emph {et~al.}(2002)\citenamefont
  {Kamiura}, \citenamefont {Hasegawa}, \citenamefont {Sano}, \citenamefont
  {Mizokawa},\ and\ \citenamefont {Kawamoto}}]{Kamiura}%
  \BibitemOpen
  \bibfield  {author} {\bibinfo {author} {\bibfnamefont {Y.}~\bibnamefont
  {Kamiura}}, \bibinfo {author} {\bibfnamefont {K.}~\bibnamefont {Hasegawa}},
  \bibinfo {author} {\bibfnamefont {Y.}~\bibnamefont {Sano}}, \bibinfo {author}
  {\bibfnamefont {Y.}~\bibnamefont {Mizokawa}}, \ and\ \bibinfo {author}
  {\bibfnamefont {K.}~\bibnamefont {Kawamoto}},\ }\href@noop {} {\bibfield
  {journal} {\bibinfo  {journal} {J. Vac. Sci. Technol.},\ }\textbf {\bibinfo
  {volume} {B 20}},\ \bibinfo {pages} {2187} (\bibinfo {year}
  {2002})}\BibitemShut {NoStop}%
\bibitem [{\citenamefont {{W.B. Ying}}\ \emph {et~al.}(1999)\citenamefont
  {{W.B. Ying}}, \citenamefont {{Y. Mizokawa}}, \citenamefont {{K. Tanahashi}},
  \citenamefont {{Y. Kamiura}}, \citenamefont {{M. Iida}}, \citenamefont {{K.
  Kawamoto}},\ and\ \citenamefont {{W.Y. Yang}}}]{Ying}%
  \BibitemOpen
  \bibfield  {author} {\bibinfo {author} {\bibnamefont {{W.B. Ying}}}, \bibinfo
  {author} {\bibnamefont {{Y. Mizokawa}}}, \bibinfo {author} {\bibnamefont {{K.
  Tanahashi}}}, \bibinfo {author} {\bibnamefont {{Y. Kamiura}}}, \bibinfo
  {author} {\bibnamefont {{M. Iida}}}, \bibinfo {author} {\bibnamefont {{K.
  Kawamoto}}}, \ and\ \bibinfo {author} {\bibnamefont {{W.Y. Yang}}},\
  }\href@noop {} {\bibfield  {journal} {\bibinfo  {journal} {Thin Solid
  Films},\ }\textbf {\bibinfo {volume} {343/344}},\ \bibinfo {pages} {393}
  (\bibinfo {year} {1999})}\BibitemShut {NoStop}%
\bibitem [{\citenamefont {Adles}\ and\ \citenamefont {Aspnes}(2008)}]{Adles}%
  \BibitemOpen
  \bibfield  {author} {\bibinfo {author} {\bibfnamefont {E.~J.}\ \bibnamefont
  {Adles}}\ and\ \bibinfo {author} {\bibfnamefont {D.~E.}\ \bibnamefont
  {Aspnes}},\ }\href@noop {} {\bibfield  {journal} {\bibinfo  {journal} {Phys.\
  Rev.\ B},\ }\textbf {\bibinfo {volume} {77}},\ \bibinfo {pages} {165102}
  (\bibinfo {year} {2008})}\BibitemShut {NoStop}%
\bibitem [{\citenamefont {{G.D. Powell}}\ \emph {et~al.}(2002)\citenamefont
  {{G.D. Powell}}, \citenamefont {{J.F. Wang}},\ and\ \citenamefont {{D.E.
  Aspnes}}}]{Powell}%
  \BibitemOpen
  \bibfield  {author} {\bibinfo {author} {\bibnamefont {{G.D. Powell}}},
  \bibinfo {author} {\bibnamefont {{J.F. Wang}}}, \ and\ \bibinfo {author}
  {\bibnamefont {{D.E. Aspnes}}},\ }\href@noop {} {\bibfield  {journal}
  {\bibinfo  {journal} {Phys.\ Rev.\ B},\ }\textbf {\bibinfo {volume} {60}},\
  \bibinfo {pages} {205320} (\bibinfo {year} {2002})}\BibitemShut {NoStop}%
\bibitem [{\citenamefont {{B. Gokce}}\ \emph {et~al.}(2010)\citenamefont {{B.
  Gokce}}, \citenamefont {{E.J. Adles}}, \citenamefont {{D.E. Aspnes}},\ and\
  \citenamefont {{K. Gundogdu}}}]{Gokce}%
  \BibitemOpen
  \bibfield  {author} {\bibinfo {author} {\bibnamefont {{B. Gokce}}}, \bibinfo
  {author} {\bibnamefont {{E.J. Adles}}}, \bibinfo {author} {\bibnamefont
  {{D.E. Aspnes}}}, \ and\ \bibinfo {author} {\bibnamefont {{K. Gundogdu}}},\
  }\href@noop {} {\bibfield  {journal} {\bibinfo  {journal} {PNAS},\ }\textbf
  {\bibinfo {volume} {107}},\ \bibinfo {pages} {17503} (\bibinfo {year}
  {2010})}\BibitemShut {NoStop}%
\bibitem [{\citenamefont {{C.P. Wade}}\ and\ \citenamefont {{C.E.D
  Chidsey}}(1997)}]{Wade}%
  \BibitemOpen
  \bibfield  {author} {\bibinfo {author} {\bibnamefont {{C.P. Wade}}}\ and\
  \bibinfo {author} {\bibnamefont {{C.E.D Chidsey}}},\ }\href@noop {}
  {\bibfield  {journal} {\bibinfo  {journal} {Appl. Phys. Lett.},\ }\textbf
  {\bibinfo {volume} {71}},\ \bibinfo {pages} {1679} (\bibinfo {year}
  {1997})}\BibitemShut {NoStop}%
\bibitem [{\citenamefont {{H. Hirayama}}\ \emph {et~al.}(1996)\citenamefont
  {{H. Hirayama}}, \citenamefont {{K. Watanabe}},\ and\ \citenamefont {{M.
  Kawada}}}]{Hirayama}%
  \BibitemOpen
  \bibfield  {author} {\bibinfo {author} {\bibnamefont {{H. Hirayama}}},
  \bibinfo {author} {\bibnamefont {{K. Watanabe}}}, \ and\ \bibinfo {author}
  {\bibnamefont {{M. Kawada}}},\ }\href@noop {} {\bibfield  {journal} {\bibinfo
   {journal} {Appl. Surf. Sci.},\ }\textbf {\bibinfo {volume} {100}},\ \bibinfo
  {pages} {460} (\bibinfo {year} {1996})}\BibitemShut {NoStop}%
\bibitem [{\citenamefont {{Z. Xu}}\ \emph {et~al.}(1997)\citenamefont {{Z.
  Xu}}, \citenamefont {{X.F. Hu}}, \citenamefont {{D. Lim}}, \citenamefont {{J.
  G. Ekerdt}},\ and\ \citenamefont {{M. C. Downer}}}]{Xu}%
  \BibitemOpen
  \bibfield  {author} {\bibinfo {author} {\bibnamefont {{Z. Xu}}}, \bibinfo
  {author} {\bibnamefont {{X.F. Hu}}}, \bibinfo {author} {\bibnamefont {{D.
  Lim}}}, \bibinfo {author} {\bibnamefont {{J. G. Ekerdt}}}, \ and\ \bibinfo
  {author} {\bibnamefont {{M. C. Downer}}},\ }\href@noop {} {\bibfield
  {journal} {\bibinfo  {journal} {J. Vac. Sci. Technol. B},\ }\textbf {\bibinfo
  {volume} {15}},\ \bibinfo {pages} {1059} (\bibinfo {year}
  {1997})}\BibitemShut {NoStop}%
\bibitem [{\citenamefont {{D.E. Aspnes}}\ and\ \citenamefont {{A.A.
  Studna}}(1979)}]{Aspnes}%
  \BibitemOpen
  \bibfield  {author} {\bibinfo {author} {\bibnamefont {{D.E. Aspnes}}}\ and\
  \bibinfo {author} {\bibnamefont {{A.A. Studna}}},\ }\href@noop {} {\bibfield
  {journal} {\bibinfo  {journal} {Appl. Opt.},\ }\textbf {\bibinfo {volume}
  {14}},\ \bibinfo {pages} {220} (\bibinfo {year} {1979})}\BibitemShut
  {NoStop}%
\bibitem [{\citenamefont {Lucovsky}(2010)}]{Lucovsky}%
  \BibitemOpen
  \bibfield  {author} {\bibinfo {author} {\bibfnamefont {G.}~\bibnamefont
  {Lucovsky}},\ }\href@noop {} {\bibfield  {journal} {\bibinfo  {journal}
  {Phys. Stat. Sol. A.},\ }\textbf {\bibinfo {volume} {207}},\ \bibinfo {pages}
  {631} (\bibinfo {year} {2010})}\BibitemShut {NoStop}%
\bibitem [{\citenamefont {Du}\ and\ \citenamefont {Corrales}(2005)}]{Du}%
  \BibitemOpen
  \bibfield  {author} {\bibinfo {author} {\bibfnamefont {J.}~\bibnamefont
  {Du}}\ and\ \bibinfo {author} {\bibfnamefont {L.~R.}\ \bibnamefont
  {Corrales}},\ }\href@noop {} {\bibfield  {journal} {\bibinfo  {journal}
  {Phys. Rev. B},\ }\textbf {\bibinfo {volume} {72}},\ \bibinfo {pages} {09221}
  (\bibinfo {year} {2005})}\BibitemShut {NoStop}%
\end{thebibliography}%

\newpage
\section{Figure Captions}
Figure 1: (a) Evolution of SHGA during air exposure of a H-terminated P-doped (111)Si sample with a carrier concentration $n=10^{12}$ cm$^{-3}$. (b) As (a) but for $n = 10^{14}$ cm$^{-3}$ . (c)  As (a) but for $n=10^{16}$ cm$^{-3}$. (d) Evolution of the average of these SHGA signals at azimuth angles of 0, 120, and 240$^{\circ}$.  Data for the As-doped sample with  $n= 10^{18}$ cm$^{-3}$ are also included.

Figure 2:  Evolutions of the hyperpolarizabilities of the up- and back bonds of these samples as calculated in the ABM.

Figure 3: Evolutions of oxide thicknesses for n-type Si as measured by SE.

\begin{figure}[h]
\includegraphics[scale = 0.8]{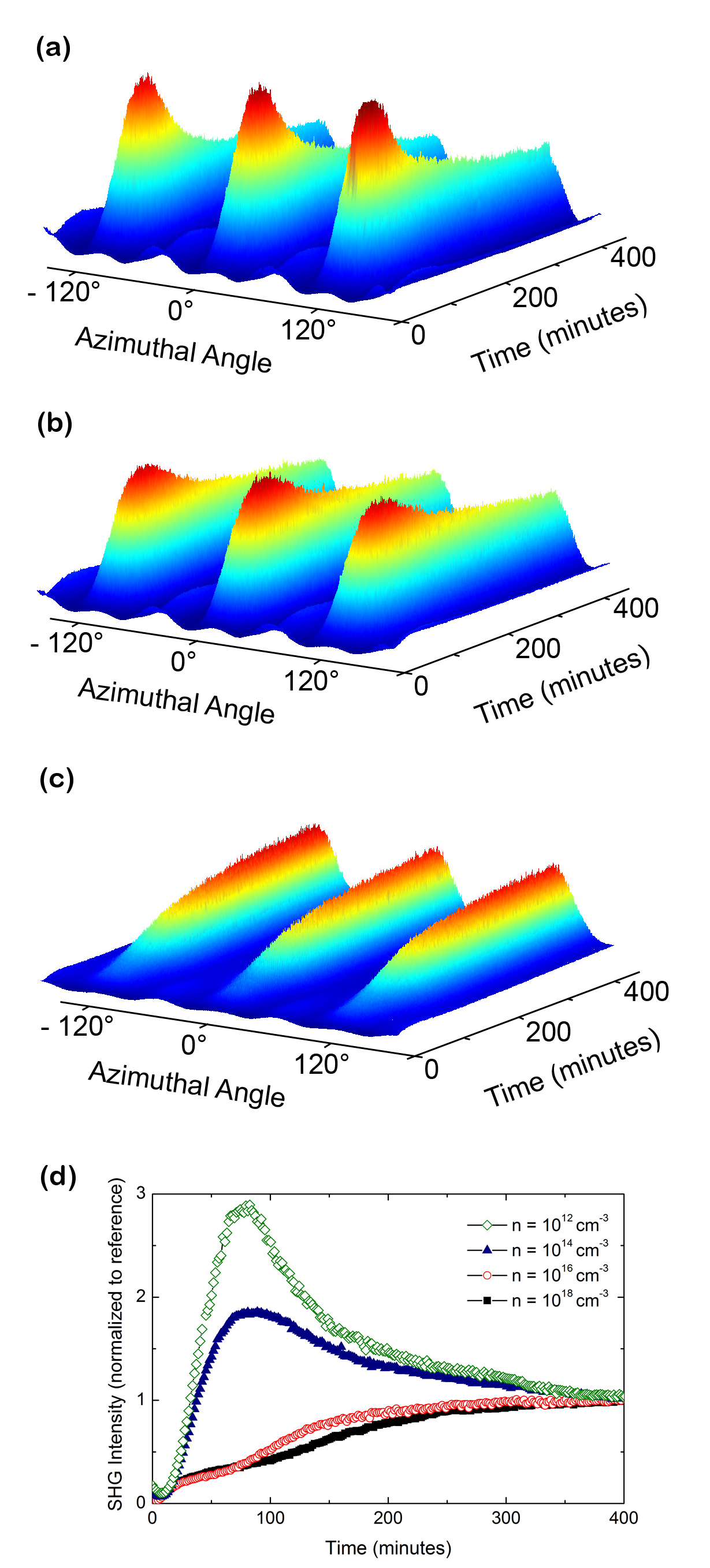}
\caption{\label{fig:epsart} }
\end{figure}

\begin{figure}[h]
\includegraphics[scale = .8]{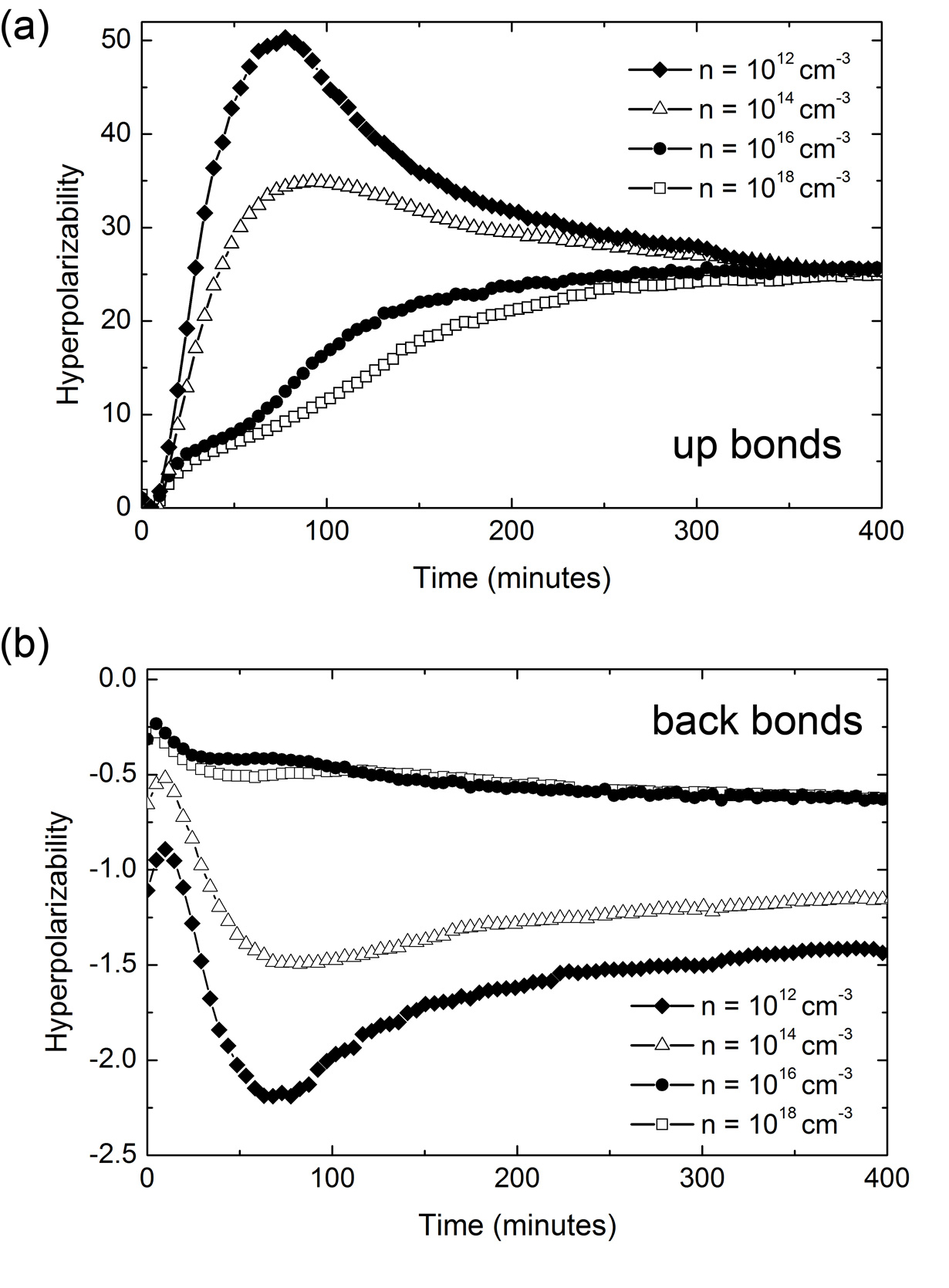}
\caption{\label{fig:epsart2} }
\end{figure}

\begin{figure}[h]
\includegraphics[scale = .8]{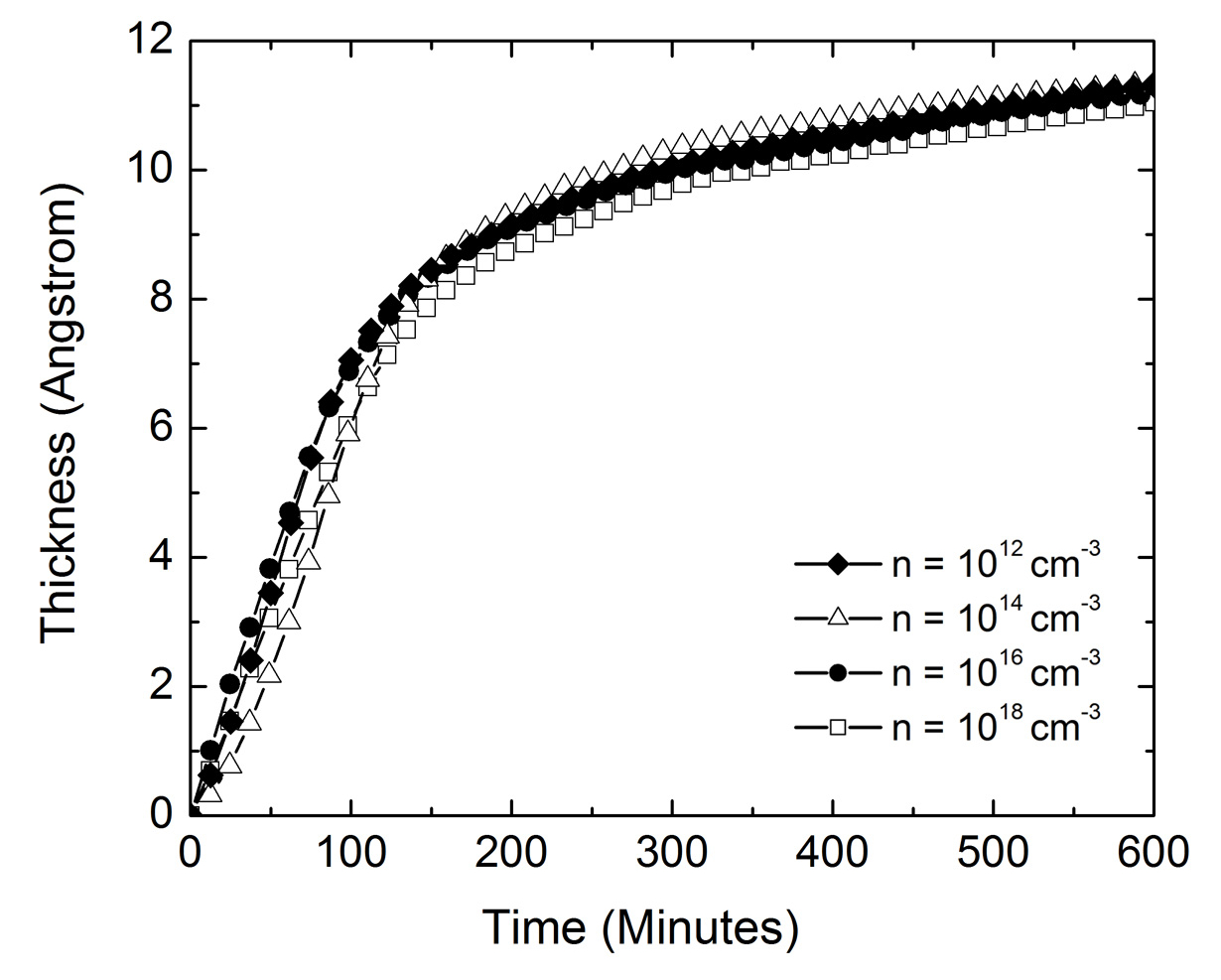}
\caption{\label{fig:epsart3}  }
\end{figure}

\end{document}